\documentclass[preprint,12pt]{elsarticle}
\usepackage{graphicx}

\usepackage{algpseudocode}
 \usepackage{algorithm}
\usepackage{amsmath}
\usepackage{amsthm}
\usepackage{xcolor}
\usepackage{makecell}
\usepackage{comment}
\usepackage{amsfonts}
\usepackage{amssymb} 
\usepackage{graphicx}
\usepackage{subcaption}
\usepackage{natbib}
\usepackage{geometry}
\theoremstyle{remark}
\newtheorem{remark}{Remark}
\usepackage[hidelinks]{hyperref}

\title{An Approach for Optimizing Acceleration in Connected and Automated Vehicles}

\author{%
  \textbf{Filippos Tzortzoglou}\\
  Cornell University, Ithaca, NY\\
  ft253@cornell.edu\\
  \hfill\break% this is a way to add line numbering on empty line
  \textbf{Dionysios Theodosis}\\
    Technical University of Crete, Chania, Greece \\
  dtheodosis@tuc.gr\\
  \hfill\break%
  \textbf{Andreas Malikopoulos}\\
  Cornell University, Ithaca, NY\\
  amaliko@cornell.edu\\
  \hfill\break%
}

\begin{document}
\makeatletter
\def\ps@pprintTitle{%
 \let\@oddhead\@empty
 \let\@evenhead\@empty
 \def\@oddfoot{}%
 \let\@evenfoot\@oddfoot}
\makeatother
\maketitle
\section{Abstract}
Vehicle automation technology has made significant progress, laying the groundwork for a future of fully automated vehicles. This paper delves into the operation of connected and automated vehicles (CAVs). In prior work, we  developed a controller that includes a tunable gain whose value significantly influences CAV performance and, in particular, its acceleration. By varying this gain, CAV acceleration is associated with different values depending on some initial conditions.
Thus, our goal in this paper is to identify the optimal value of this gain in terms of acceleration for any group of initial conditions. To this end, we formulate an optimization problem where the decision variable is the gain value, and the objective function includes the acceleration of the vehicles. The complexity of this problem prohibits real-time solutions. To address this challenge, we train a neural network to map different initial conditions to the optimal gain values efficiently. We showcase the proposed approach to deriving the optimal gains in a merging scenario with an on-ramp.
% \noteIshti{Please include one/more sentences of the concluding remarks/main results, if available.}

\noindent\textit{Keywords}: Connected and Automated Vehicles, Optimization
\newpage

\section{Introduction}

% \noteJake{The Shladover papers and any others that we can find on other impacts analyses using microsimulation to model CACC are going to be the most important part of the introduction}

% \noteJeff{This is an example comment}

Over the last decades, we have witnessed an increase in the merging of digital networks, energy, and transportation \cite{campisi2021development, nikitas2020artificial,goldbeck2019resilience}. This merging, along with human interactions, is making these systems more complex \cite{schadschneider2010stochastic,zeng2017science,dave2023worst}. Congestion in certain areas of the US has led to American commuters experiencing a significant increase in travel time that has subsequently increased fuel consumption, with an extra 3.1 billion gallons of fuel being purchased \cite{schrank20152015}. Consequently, new control strategies are needed to manage vehicle interactions in different traffic scenarios \cite{zhu2018analysis,zhang2021design, houshmand2019penetration}. 

In recent years,  CAVs have attracted considerable attention as they seem able to address most of these issues \cite{guanetti2018control,liu2019systematic,xu2019grouping}. Their ability to share information about traffic conditions contributes to enhanced safety, reduced travel time, and improved fuel economy \cite{Rios-Torres:2017aa,malikopoulos2021optimal,li2018separation}. Numerous research initiatives have explored the application of CAVs in different traffic scenarios, such as intersections, roundabouts, merging roadways, and speed reduction zones \cite{xu2021comparison,chalaki2020experimental, venkatesh2023connected}. Similarly, adaptive cruise control (ACC) and cooperative ACC have shown that they can increase road capacity and improve traffic flow \cite{xiao2010comprehensive,wang2018review,ntousakis2015microscopic}. These approaches have demonstrated the potential to improve fuel efficiency, save energy,  and enhance throughput while ensuring safety. 

In previous work \cite{theodosis2022sampled,karafyllis2022stability}, a Control Lyapunov function methodology was employed to design a controller to optimize the flow of CAVs on both single-lane and lane-free roads. The resulting controller incorporates numerous free functions and adjustable gains, which have the potential to influence the vehicles' performance. However, the selection of these parameters was conducted arbitrarily without the use of a systematic methodology. 

In this paper, we derive the optimal value for the gain embedded in the controller presented in \cite{karafyllis2022stability}, aiming to minimize the acceleration of the vehicles, improving energy efficiency and enhancing passenger comfort.  We consider scenarios involving vehicles operating on single-lane roads, as in \cite{karafyllis2022stability} while examining a case study that involves a merging scenario via an on-ramp. 

 The optimal gain value is not fixed as it is significantly influenced by the other vehicles' initial conditions (speed and position) interacting with the CAV. Thus, varying these initial conditions would lead to different acceleration profiles for the CAV.  Consequently, we aim to find the optimal value of this gain for different initial conditions via an optimization problem where the objective function includes the acceleration of the CAV, and the value of the gain constitutes the decision variable. Given the significant challenges associated with real-time numerical solutions for this optimization problem, we have utilized a neural network to solve it. After generating thousands of data points offline, we learn the correlation between the initial conditions and the optimal gain values. Simulation results showed the effectiveness of this approach in obtaining smaller accelerations/decelerations compared to \cite{karafyllis2022stability}.

To underline the importance of finding the optimal gain according to the initial conditions,  we enhance our controller's operation to handle a merging scenario via an on-ramp. Similar to \cite{7534837}, we use a coordinator that manages the merge queue and assigns a new gain to the vehicles when a new vehicle enters the main road from an on-ramp. 

The remainder of the paper is organized as follows. In Section 2, we present our problem formulation. Section 3 is dedicated to demonstrating our solution approach. In Section 4, we provide results that confirm the effectiveness of our approach, while Section 5 extends our approach to a merging scenario. Finally, we provide some concluding remarks in Section 6.

\section{Problem Formulation}
\begin{figure}[thp]
\centering
\includegraphics[scale=0.45]{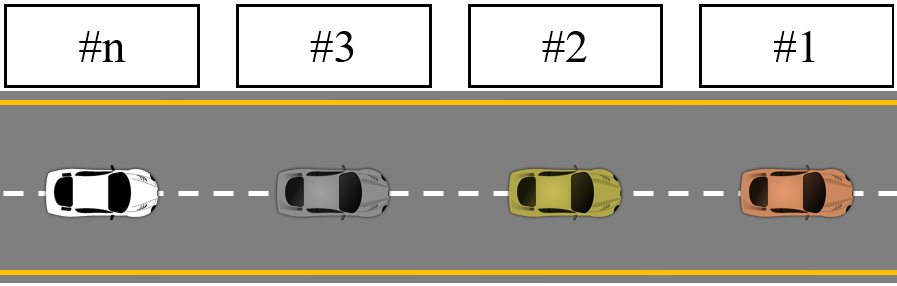}
\caption{Single lane road with four vehicles}
\label{fig:Lane_1}
\end{figure}
Consider $n\in\mathbb{N}$ vehicles moving on a single-lane road, as shown in FIGURE \ref{fig:Lane_1}. The movement of these vehicles is defined by:

\begin{equation} \label{eq:initial model}
    \begin{aligned}
      &\dot{x}_i =  v_i,\,\,\, i=1,\ldots,n,\\
      &\dot{v}_1 = F_1(v_1,s_2), \\
      & \dot{v}_i = F_i(v_i,s_i,s_{i+1}), \,\,i=2,\ldots,n-1, \\
      &\dot{v}_n = F_n(v_n,s_n),
          \end{aligned}
  \end{equation}

\noindent where  $s_i:=x_{i-1}-x_i$, $i=2,\ldots,n$ describes the inter-vehicle distance (back-to-back distance), $x_i$ is the position, $v_i$ is the speed, and $\dot{v}_i$ is the acceleration of the $i$-th vehicle. The functions $F_i$, $i=2,\ldots,n$, define the bidirectional feedback laws (cruise controllers) as defined in \cite{karafyllis2022stability}  (for brevity, the arguments of $F_i$ have been omitted):

  \begin{equation} \label{eq:initial model1}
    \begin{aligned}
      &{F}_1 = -k_1(s_2)(v_1 - v^*) - V'(s_2), \\
      &{F}_i= -k_i(s_i, s_{i+1})(v_i - v^*) + V'(s_i) - V'(s_{i+1}), \\
      &{F}_n = -k_n(s_n)(v_n - v^*) + V'(s_n),
          \end{aligned}
  \end{equation}

 \noindent where the terms $k_i$ represent state-dependent gains ensuring that the velocity of each vehicle stays positive but does not exceed the speed limit $v_{max}$, and the constant $v^*\in (0,v_{\max})$ signifies the desired speed. The function $V$ is a potential function that generates a repulsive force for collision prevention. When the distance $s$ between two vehicles gets short, this function yields higher values, whereas when the distance exceeds a certain threshold, no repulsion is induced. More specifically, the function $V$ satisfies the following conditions:

\begin{equation}\label{eq:V_properties}
    \begin{aligned}
\lim_{s \to L^+} V(s) & = +\infty, \\
V(s)& =0, \quad \forall\; s\geq\lambda,        
    \end{aligned}
\end{equation}

\noindent where $L>0$ defines the minimum inter-vehicle distance, and $\lambda>L$ indicates the distance at which vehicles will no longer exert repulsive forces on each other. The terms $k_i$ are defined as:

   \begin{equation} \label{eq:initial model2}
    \begin{aligned}   
      &k_1(s_2) = \mu + g(-V'(s_2)), \\
      &k_i(s_i, s_{i+1}) = \mu + g(V'(s_i) - V'(s_{i+1})), \\
      &k_n(s_n) = \mu + g(V'(s_n)),
    \end{aligned}
  \end{equation}

\noindent where $\mu>0$ represents a gain that influences the speed at which vehicles converge to their desired speed (settling time). The function $g$ is defined by

\begin{equation} 
\label{eq:g_function}
g(x)=\frac{v_{\max}f(x)}{v^{*}(v_{\max}-v^{*})}-\frac{x}{v^*},\,\,x\in\mathbb{R},
\end{equation}

\noindent   where $f(x)$  is non-decreasing function that satisfies $\max(x,0)\leq f(x)\; \forall\,\,x \in \mathbb{R}.$ Here, we select the function $f$ as in \cite{karafyllis2022stability}:

\begin{equation}
  f(x) =\frac{1}{2\epsilon} \begin{cases}
  0, & \text{if } x \leq -\epsilon, \\
  (x+\epsilon)^2, & \text{if } -\epsilon < x < 0, \\
  \epsilon^2 +2 \epsilon x & \text{if } x \geq 0,
\end{cases}
\end{equation}

\noindent with $\epsilon>0$. The state space of the closed-loop system \eqref{eq:initial model}, \eqref{eq:initial model1} is

\begin{equation} \label{state_space} 
\begin{aligned}\Omega =&\left\{\, (s_{2} ,...,s_{n} ,v_{1} ,...,v_{n} )\in \mathbb{R} ^{2n-1} \, :\, \mathop{\min }\limits_{i=2,...,n} (s_{i} )>L\, ,\right.\left. \mathop{\max }\limits_{i=1,...,n} (v_{i} )\le v_{\max }, 
\mathop{\min }\limits_{i=1,\ldots,n} (v_{i} )\ge 0\, \right\}.  \end{aligned}
\end{equation} 

\noindent Equations \eqref{eq:initial model1} guarantee that for any initial condition $(s_2(0),\ldots,s_n(0),v_1(0),$ $\ldots,v_n(0))\in \Omega$ the unique solution of \eqref{eq:initial model}, \eqref{eq:initial model1} satisfies $(s_2(t),\ldots,s_n(t),v_1(t),\ldots,v_n(t))\in \Omega$  for all $t\ge 0$ and $\lim_{t\to+\infty}$ $v_i(t)=v^*$ for all $i=1,\ldots,n$.

\subsection{Influence of the gain $\mu$}
In \eqref{eq:initial model2}, we observe the gain $\mu$ that constitutes a term of the state-dependent gains $k_i$, $i=1,\ldots,n$, which are directly connected with our feedback laws $F_i$ defined in \eqref{eq:initial model1}. The value of the gain $\mu$ plays a critical role since it influences the time it takes for vehicles to converge to their desired speed. Namely, when the gain $\mu$ is set to a high value, vehicles will converge to their desired speeds faster, leading to higher accelerations. Conversely, a lower value would result in reduced vehicle accelerations, thus extending the time for convergence of the speeds. However, while it might seem reasonable to set the value of this gain as small as possible, simulation results have shown that this approach is not optimal in terms of acceleration.  In prior work \cite{theodosis2022sampled,karafyllis2022stability}, a fixed gain value of $\mu=0.5$ was selected which might not be optimal.

\section{Solution}
In the following section, we delve into our approach for determining the optimal value of the gain, $\mu$, in terms of acceleration. To achieve this, we formulate the following optimization problem:

\begin{equation} \label{optimization problem}
\begin{aligned}
\min_{\mu} \quad \int_{t_s}^{t_e}& \sum_{i=1}^{n} \dot{v}_i^2\; dt   \\
\text{subject to} \quad 0 &< \mu \leq 2,  \\
\dot{v}_{min}&\leq \dot{v}_i \leq \dot{v}_{max}. 
\end{aligned}
\end{equation}

The objective function in \eqref{optimization problem} is constituted by one term which pertains to the vehicles' acceleration. The variables $t_s$ and $t_e$ correspond to the time horizon within which we aim to solve the optimization problem, and the gain $\mu$ is the decision variable. The feasibility domain of the decision variable $\mu$ has been chosen to span the interval $(0,2]$. According to \cite{karafyllis2022stability}, while the gain $\mu$ is required to be positive, no upper bound is imposed. Nonetheless, we have chosen to set the number 2 as the upper bound since when the value of $\mu$ exceeds 2, the controller becomes excessively sensitive, resulting in high acceleration even with minor deviations from the desired speed. Finally, we impose restrictions on the vehicle's accelerations and decelerations, ensuring they do not exceed two predefined limits $\dot{v}_{max}$ and $\dot{v}_{min}$. Here, we select these limits as 3.5 m/s² and -4 m/s², respectively.

\begin{remark}
Note that the dynamics of the vehicles in ($\ref{eq:initial model}$) constitute an initial value problem. Hence the evolution of the accelerations is influenced by the initial conditions of the vehicles \cite{karafyllis2022stability}. This finding indicates that we need to solve a different optimization problem \eqref{optimization problem} every time we deal with different initial conditions. The importance of investigating various initial conditions and the corresponding optimal gain, $\mu$, lies in understanding the dynamic characteristics of the traffic flow. More specifically, when a new vehicle merges onto a single-lane road, it is necessary to address the initial value problem outlined in equation \eqref{eq:initial model} to determine the updated trajectories of all vehicles. In other words, we need to resolve the same system of ordinary differential equations, but the initial conditions now incorporate the newly added vehicle.
\end{remark}

\begin{remark}
Note that if the initial speed deviates considerably from the desired speed, the optimization problem might not have a feasible solution since the second constraint can be violated regardless of the value of the gain $\mu$. As illustrated in \eqref{eq:initial model1}, the deviation from the desired speed proportionally influences the magnitude of the first term of $F_i$ which is equal to $-k_i(s_i, s_{i+1})(v_i - v^*)$, possibly leading to high values of $F_i$. Namely, high deviations from the desired speed can result in unrealistic accelerations. Consequently, although tuning the parameter $\mu$ can result in better acceleration profiles, it cannot guarantee that the resulting accelerations will fall within the interval [$\dot{v}_{min}$, $\dot{v}_{max}$]. Ongoing research is focused on identifying specific families of initial conditions that guarantee bounded accelerations.  
\end{remark}

\subsection{Solution approach}
Obtaining an analytical solution to the optimization problem \eqref{optimization problem} is nontrivial. From \eqref{eq:initial model}, \eqref{eq:initial model1}, and \eqref{eq:initial model2}, we note that our feedback laws are nonlinear. The complexity of these feedback laws and the intricate interactions among the functions $V$, $g$, and $f$ preclude such an analytical approach. To circumvent these challenges, we propose to approximate the solution of \eqref{optimization problem} using numerical methods. While these methods can provide highly accurate approximations of the optimal solution, they also require substantial computational time, which may not be feasible for real-time applications.

\subsection{Implementation of neural network}
\begin{figure}
\centering
\includegraphics[width=0.8\textwidth]{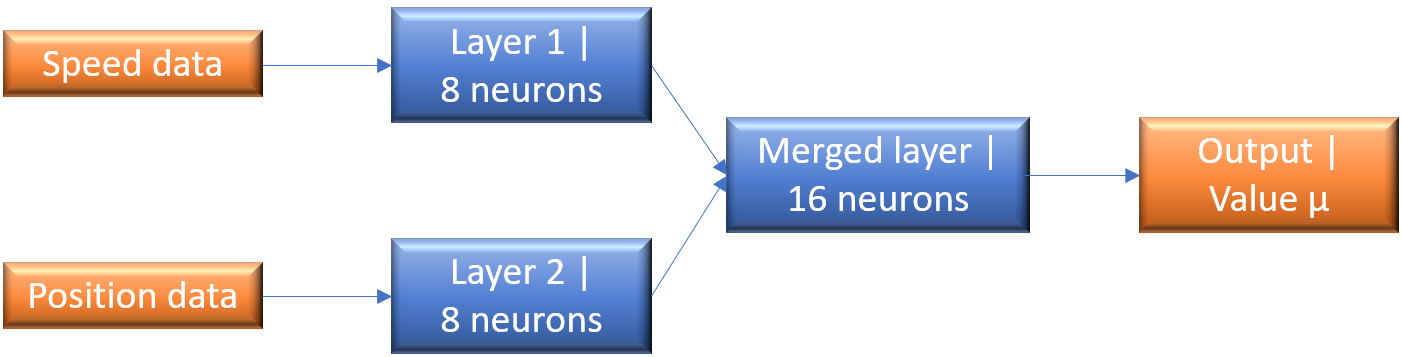}
\caption{Neural network architecture.}
\label{fig:NN_Arxchitecture}
\end{figure}

To accommodate real-time implementation, we utilize a neural network to learn the relationship between the initial conditions of the system and the related efficient value of the gain $\mu$. To train the neural network, we generate a data collection consisting of varying initial conditions and corresponding solutions to the optimization problem \eqref{optimization problem}. In order to precisely capture the different roles of initial speeds and positions (i.e., initial conditions) in our system, we implement a unique architecture for our neural network as illustrated in FIGURE \ref{fig:NN_Arxchitecture}. The network is structured to ensure initial non-overlap between the inputs representing these two data. For example, in the case of seven vehicles, we feed the input layer of the network with fourteen elements, composed of the seven-speed elements and seven-position elements for each of the seven vehicles in our system. However, these elements do not directly interact. Instead, the network initially processes speed and position data separately, to reflect their individual interpretations in the first and second layers. Following this independent initial processing, the subsequent layer gradually integrates the learned information from both previous layers, thus enabling the model to generate predictions that account for the interconnectedness between the positions and speeds of the vehicles.

\begin{figure}
\centering
\includegraphics[width=0.6\textwidth]{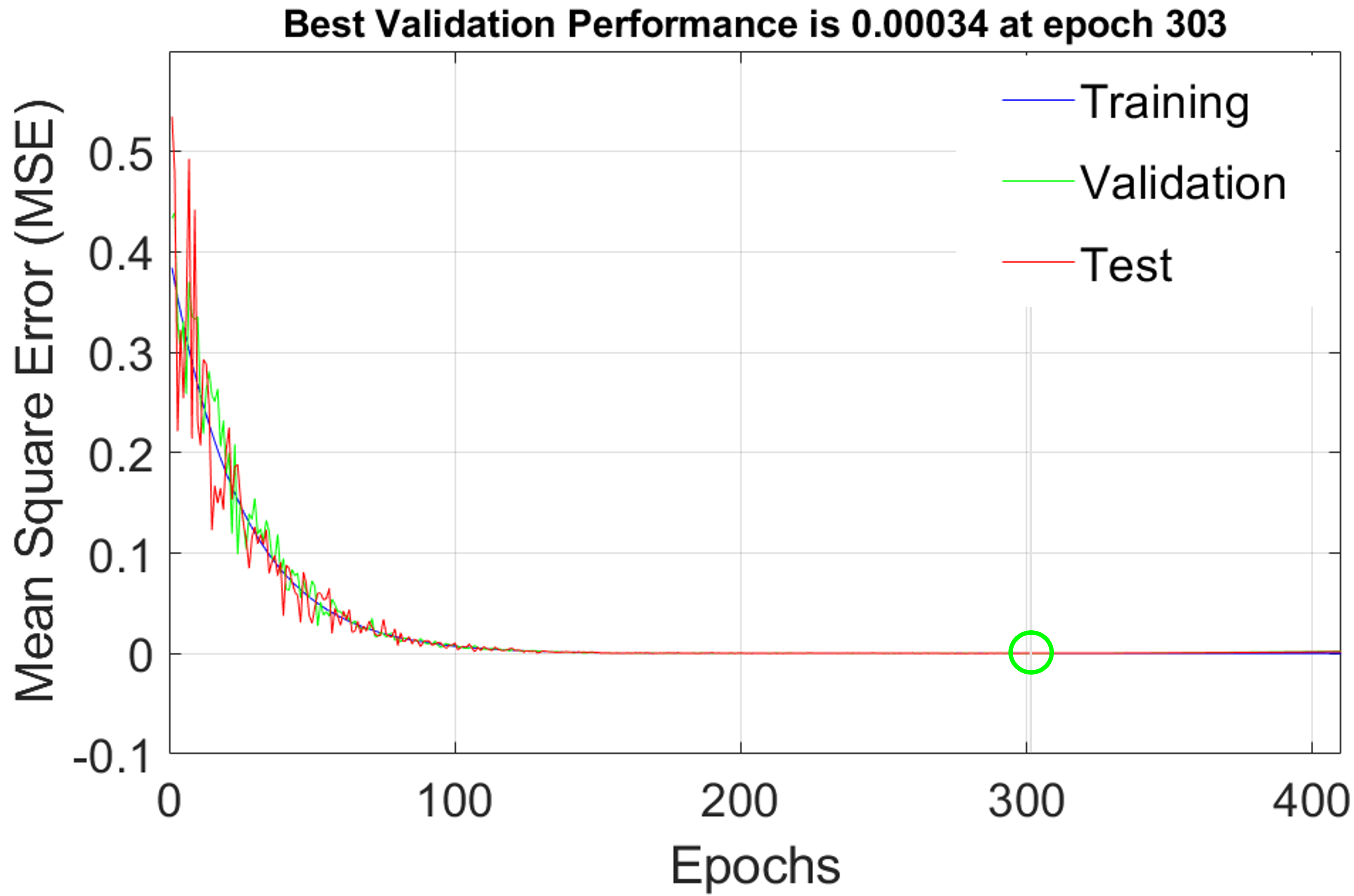}
\caption{Performance of the neural network.}
\label{fig:Neural Net}
\end{figure}

To facilitate the training of the neural network we normalized our data points. We generated 5,000 data points and split them as follows: 85$\%$ for training, 7.5$\%$ for validation, $7.5\;\%$ for testing. Each neuron's output underwent a rectified linear unit (ReLU) activation function. Training of the network used the backpropagation method with a mean squared error (MSE) loss function and a learning rate of 0.00075. The training endured until it had run through 400 epochs. The network's final MSE was 0.00034 at epoch 303 which indicates that, on average, each resulting value $\mu$ deviated from the actual one (provided by the numerical solution) by $0.0184$. In FIGURE \ref{fig:Neural Net}, we see the training process of the neural network.

After the network is trained, it can rapidly predict the efficient value of the gain $\mu$ for any given set of initial conditions. This method significantly lessens the computational burden during real-time applications without sacrificing the quality of the solution. 

\begin{remark}
The construction of the neural network was implemented using seven vehicles to demonstrate the effectiveness of our approach. However, our final analysis is not limited to this number and can accommodate any quantity of vehicles.  
\end{remark}

\begin{remark}
To ensure realistic initial conditions of the vehicles we restrict the initial inter-vehicle distances, to be greater than or equal to the value $s_i=\overline{s}+\rho v_{i}$ where $\overline{s}$ is a standstill distance,  $\rho$ is the minimum headway that the rear vehicle maintains following the preceding CAV and $v_{i}$ is the initial speed of the rear CAV.
\end{remark}

 \subsection{Results}

In this section, we compare our approach with the approach presented in \cite{karafyllis2022stability}. In both cases, we utilized fixed initial conditions to effectively demonstrate the advancements in acceleration achieved through the appropriate tuning of the gain $\mu$. We consider $n=7$ vehicles with initial spacing $s_i(0)\;\in\;(16,24)$ and initial speed $v_i(0)\;\in\;(27,34)$. We select the parameters of the controllers as selected in \cite{karafyllis2022stability}, that is: $L=5$ m, $\lambda=20$ m, $v^*=30$ m/s, $v_{max}=35$ m/s, $\epsilon=0.2$. Also, let

\begin{equation}
    V(s) = 
    \begin{cases} 
         \frac{  (\lambda - s)^3 }{(s - L)}, & \text{if } L < s < \lambda,\\
        0, & \text{if } \lambda \geq s.
    \end{cases}\\
    \label{new_type_of_potential}   
\end{equation}

\begin{figure*}[h!]
    \centering
    
    \begin{subfigure}[b]{0.48\textwidth}
        \includegraphics[width=\textwidth]{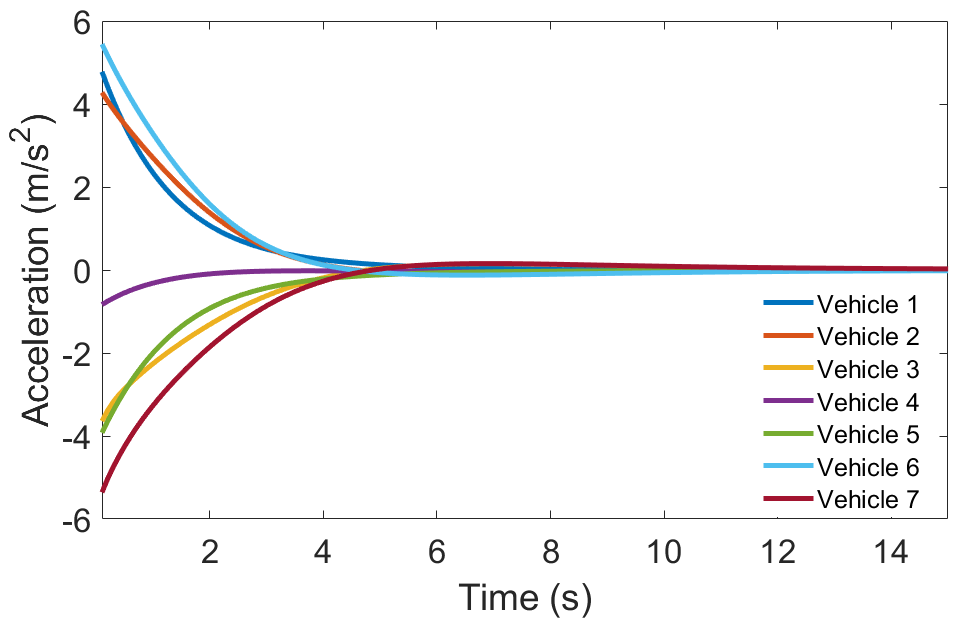}
        \caption{Acceleration with $\mu$ as in \cite{karafyllis2022stability}.}
        \label{fig:Acceleration_old_multi}

    \end{subfigure}
    \hspace{1.5mm}
    \begin{subfigure}[b]{0.48\textwidth}
        \includegraphics[width=\textwidth]{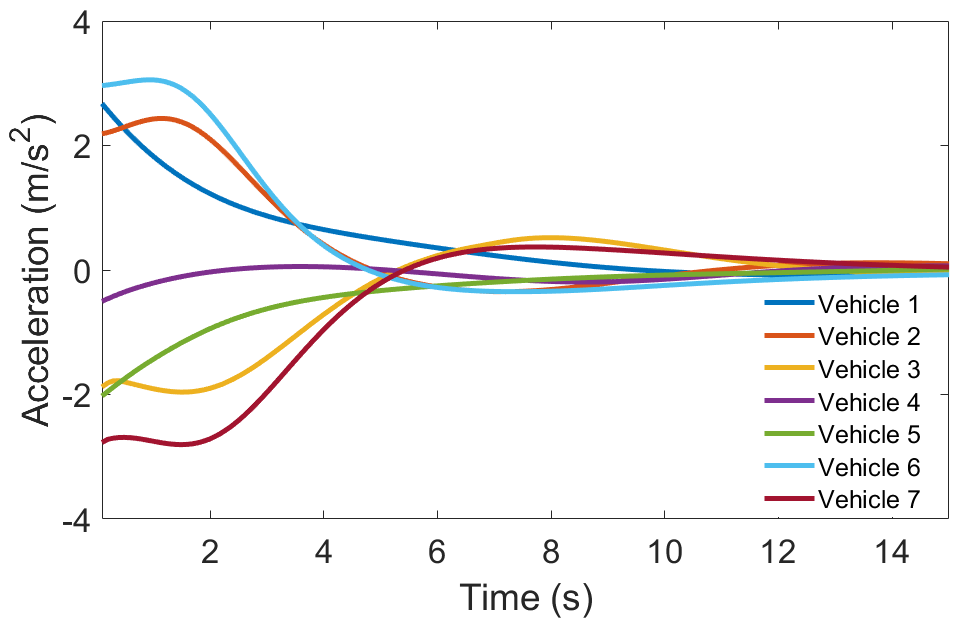}
        \caption{Acceleration using our new approach.}
        \label{fig:Acceleration_new_multi} 
    
    \end{subfigure}
        \caption{Comparison of accelerations using the approach presented in \cite{karafyllis2022stability} and the proposed approach.}
    \label{fig:Accelerations_multi}

\end{figure*}

\begin{figure*}[h!]
    \centering
    
    \begin{subfigure}[b]{0.48\textwidth}
        \includegraphics[width=\textwidth]{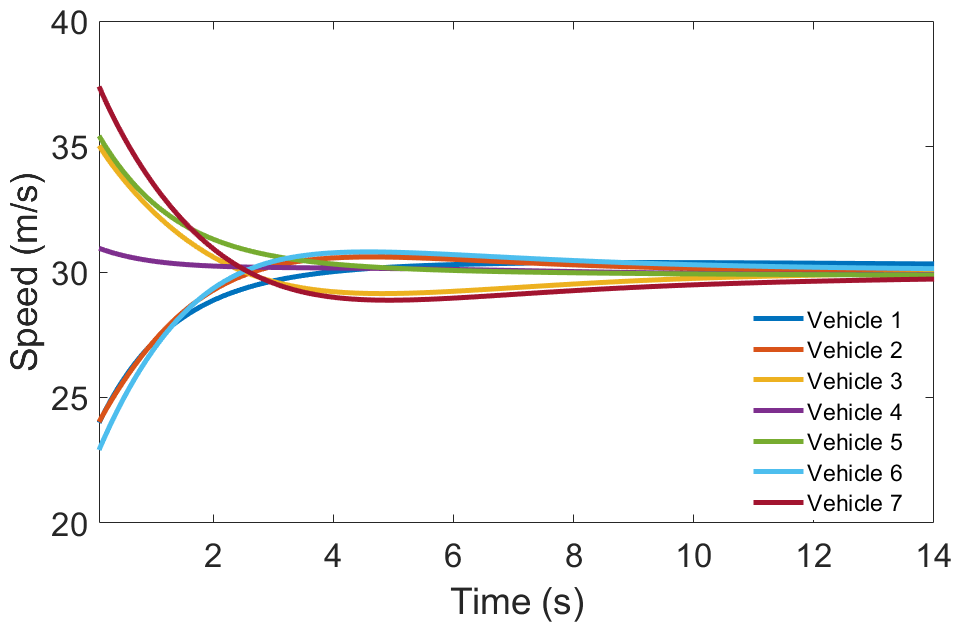}
        \caption{Speed with $\mu$ as in \cite{karafyllis2022stability}.}
        \label{fig:Speed_old}

    \end{subfigure}
    \hspace{1.5mm}
    \begin{subfigure}[b]{0.48\textwidth}
        \includegraphics[width=\textwidth]{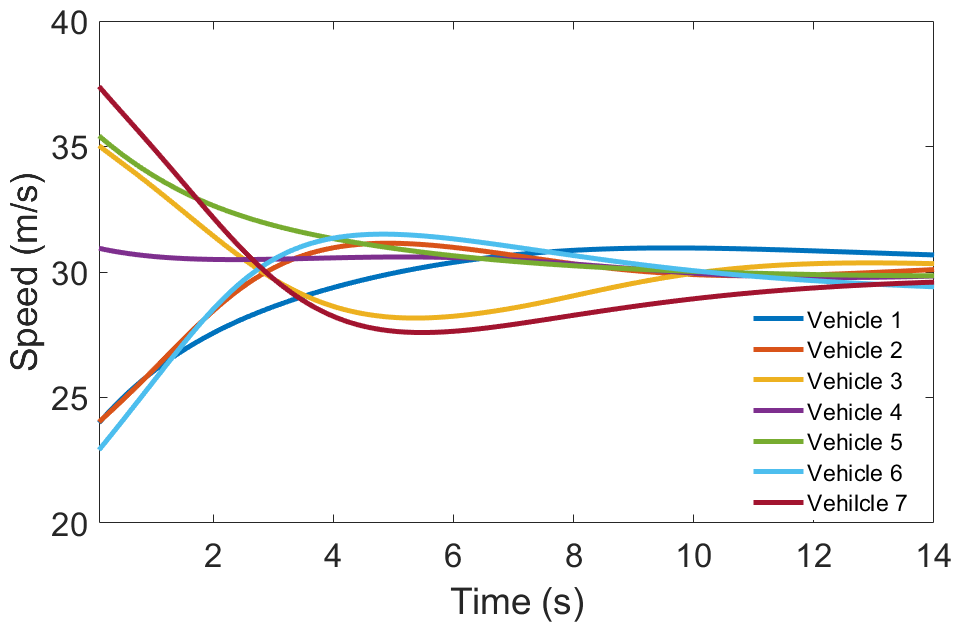}
        \caption{Speed using our new approach.}
        \label{fig:Speed_new} 
    
    \end{subfigure}
        \caption{Comparison of speeds using the approach presented in \cite{karafyllis2022stability} and the proposed approach.}
    \label{fig:speeds_multi}

\end{figure*}

In FIGURE \ref{fig:Acceleration_old_multi}, the acceleration of the vehicles is shown, considering the gain $\mu$ equal to 0.5, as defined in \cite{karafyllis2022stability}. It is noticeable that vehicles 6 and 7 exhibit initial acceleration and deceleration values that are rather unrealistic, exceeding $\pm 5$ m/s². Also it is notable that the accelerations tend to converge to 0 m/s² after the first 5 seconds something that can lead to an undesirable driving experience. FIGURE \ref{fig:Acceleration_new_multi} displays the acceleration of the vehicles under our newly proposed method. We observe that our approach significantly reduces the maximum acceleration/deceleration to $\pm$3 m/s² indicating a significant improvement higher than to 40$\%$. Moreover, the difference in the rate of convergence of the speeds can be seen in FIGURES \ref{fig:Speed_old} and \ref{fig:Speed_new}, which represent the old and new approach, respectively. We confirm in FIGURE \ref{fig:Speed_new} that the speeds tend to converge to the desired speed more slowly compared to FIGURE \ref{fig:Speed_old}, thereby improving passenger comfort.

\section{Extension of our analysis in a Merging Scenario}

\begin{figure*}[h!]
    \centering
    
    \begin{subfigure}[b]{0.45\textwidth}
        \includegraphics[width=\textwidth]{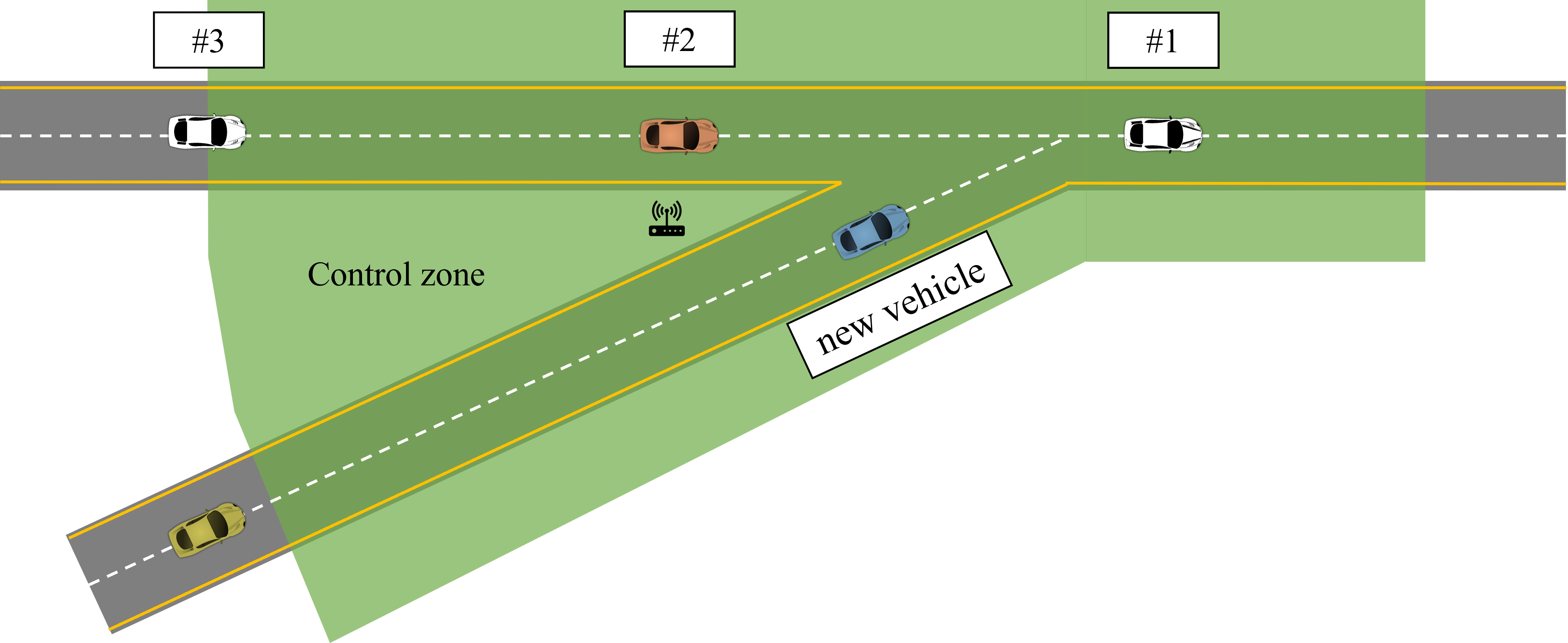}
        \caption{Snapshot 1}
        \label{fig:snapshot 1}

    \end{subfigure}
    \hspace{1.5mm}
    \begin{subfigure}[b]{0.45\textwidth}
        \includegraphics[width=\textwidth]{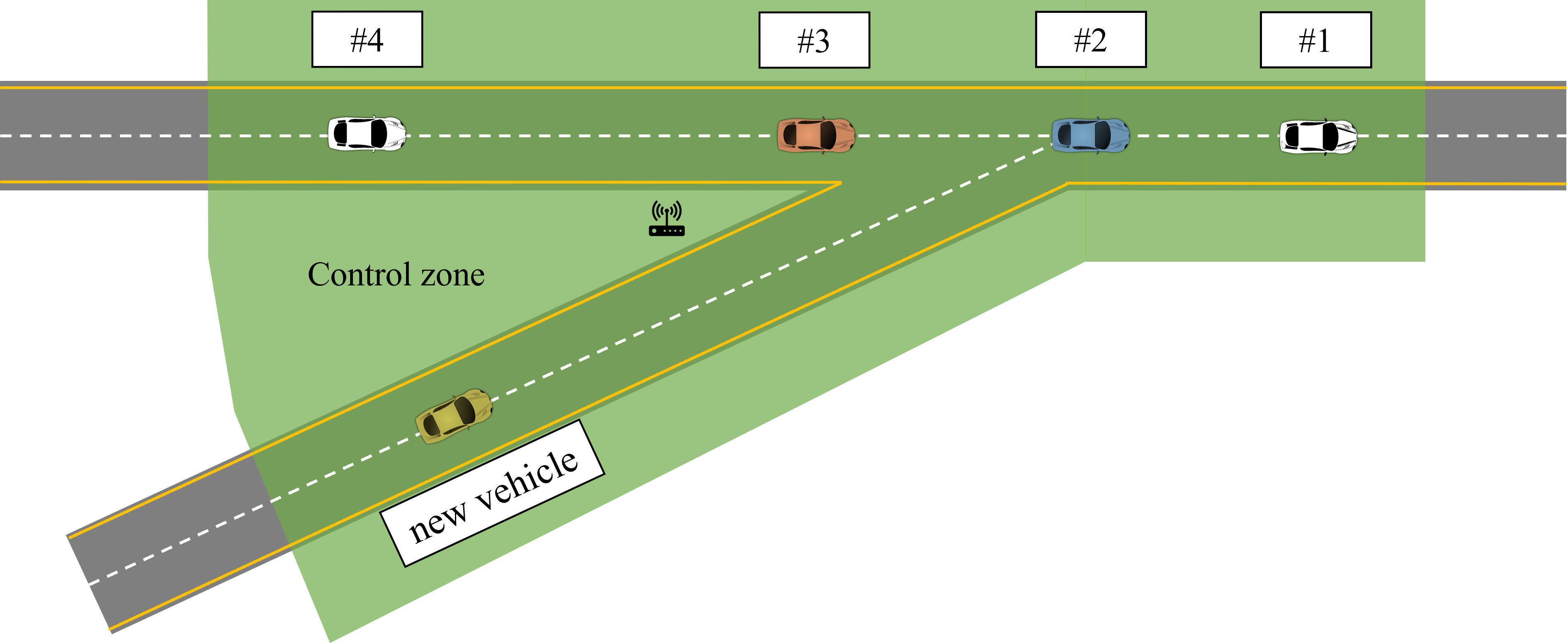}
        \caption{Snapshot 2}
        \label{fig:snapshot 2} 
    
    \end{subfigure}
        \caption{Two consecutive snapshots of a merging scenario via on-ramp.}
    \label{fig:Snapshots}

\end{figure*}

This section aims to showcase vehicles' ability to achieve smaller accelerations under various initial conditions. Our analysis is applicable when an incoming vehicle joins the road (e.g., from an on-ramp). Each time a new vehicle enters the main road, we resolve the initial value problem outlined in \eqref{eq:initial model}, incorporating the incoming vehicle into the initial conditions. For instance, in FIGURE \ref{fig:snapshot 1}, we see a new vehicle that is approaching the main road. In FIGURE \ref{fig:snapshot 2}, the same vehicle is shown to have just merged into the main road. At this specific time, we need to resolve the initial value problem defined in \eqref{eq:initial model} considering the new vehicle. Next, we demonstrate the effectiveness of our approach in the context of a merging scenario involving a primary and a secondary road, as illustrated in FIGURE \ref{fig:Snapshots}.  A coordinator manages the assignment of the new gain $\mu$ to the vehicles. The coordinator has access to the positions and speeds of the vehicles inside a control zone, enabling it to rapidly allocate the most efficient gain value $\mu$ to the system when a new vehicle merges into the main road.

The process by which vehicles from the secondary road join the main road is determined using the framework presented in \cite{7534837}. Let $MR=\{1,2,...n\}$ be the set of vehicles in the main road and $SR=\{1,2,...m\}$  the set of vehicles in the secondary road. The trajectories of the acceleration (control input), speed, and position for each vehicle $i \in SR$ are defined as follows:

\begin{equation}
\begin{aligned}
    \dot{v}_i(t)&=\alpha_{i}t+b_i,\\
    v_i(t)&=\frac{1}{2}\alpha_it^2+b_it+c_i,\\
    x_i(t)&=\frac{1}{6}\alpha_it^3+\frac{1}{2}b_it^2+c_it+d_i,\\ 
\end{aligned}
\label{hamiltonian}
\end{equation}

\noindent where $\alpha_i, b_i, c_i$ and $d_i$ are constants of integration and can be found through:

\begin{equation}
\label{eq:phi_i}
\begin{bmatrix}
a_i 
\\
b_i 
\\
c_i 
\\
d_i
\end{bmatrix}
= 
\begin{bmatrix}
\frac{1}{6}t^3 & \frac{1}{2}t^2 & t & 1 
\\
\frac{1}{2}t^2 & t & 1 & 0 
\\
\frac{1}{6}(t_i^f)^3 & \frac{1}{2}(t_i^f)^2 & t_i^f & 1 
\\
\frac{1}{2}(t_i^f)^2 & t_i^f & 1 & 0 
\end{bmatrix}^{-1}
\begin{bmatrix}
x_i(t) 
\\
v_i(t) 
\\
x_i(t_i^f) 
\\
v_i(t_i^f)
\end{bmatrix}.
\end{equation}

\noindent where $t_i^f\; \forall \; i \in MR\cup SR $ is the time when the vehicles reach the point where the two roads are merged. The velocities of the vehicles at each time moment $t_i^f\;\forall\;i\;\in\;SR$ are randomly selected within the interval [23 m/s, $v_{max}$]. This is done to demonstrate the implementability of our approach under a variety of initial speeds. Recall that every time a new vehicle merges into the main road we resolve the initial value problem defined in (\ref{eq:initial model}).  

To prevent a potential conflict between a CAV $i\; \in MN$ and a CAV $j \; \in SR$ we require a minimum time gap ${t_{\min}>0}$ between the time instants ${t_i^f}$ and $t_j^f$ when the CAV $i$ and CAV $j$ cross the point of conflict of the two roads. That is:

\begin{equation}\label{eq:lateral_constraint}
|t_i^f - t_j^f| \geq t_{min}, \;\;\;\forall i\in MN,\;\;\; \forall j\in SR.
\end{equation}

\begin{remark}
Note that the range of the coordinator is limited and cannot encompass every vehicle on the main road. As a result, whenever a new vehicle merges, the coordinator assigns the new gain $\mu$ exclusively to vehicles within the control zone. This limitation does not hinder our analysis since different vehicles might possess varying values of the gain $\mu$ \cite{karafyllis2022stability}.
\end{remark}

\begin{remark}
We use the controllers presented in \cite{7534837} only for the vehicles operating on the secondary road. By the time the vehicles from the secondary road merge into the main road, they adapt the controller framework presented in (\ref{eq:initial model}). 
\end{remark}

\begin{remark}
In this section, our goal is to showcase that our approach is real-time implementable under various initial conditions. We borrow the controllers from \cite{7534837} just to safely merge vehicles from the secondary road into the main road while minimizing the control input. \\
\\ It is clear that the selection of $t_i^f\;\forall\;i\;\in SR$ influences the throughput of the vehicles on the secondary road. Consequently, in order to find the minimum $t_i^f\; \forall\; i \in SR$ and maximize the throughput on the secondary road we formulate the following minimization problem: 
\end{remark}

\begin{equation}
\begin{aligned}
\label{optimization problem2}
\min &\quad t_i^f \;\;\;\;\; \forall \; i \in SR \\
\text{subject to} &\quad v_{min} < v_i \leq v_{max}, \;\;\;  \\
&\quad \dot{v}_{min}\leq \dot{v}_i \leq \dot{v}_{max}, \;\;\;\;\;\;\;\; \forall \; i \in SR \\
&\quad p_{i}(t) - p_{i-1}(t) \geq  \overline{s}+\rho v_{i}.  \\
&\quad \eqref{eq:lateral_constraint} 
\end{aligned}
\end{equation}

\noindent Here, the first two constraints are related to the speed limits and the acceleration boundaries, while the $3^{rd}$ constraint ensures that there is no rear-end collision between the vehicles on the secondary road. Regarding the value $\overline{s}+\rho v_{i}$ see Remark 4. For the solution of this optimization problem, see \cite{chalaki2020experimental}.

\subsection{Simulation Setup}

\begin{figure}
\centering
\includegraphics[width=1\textwidth]{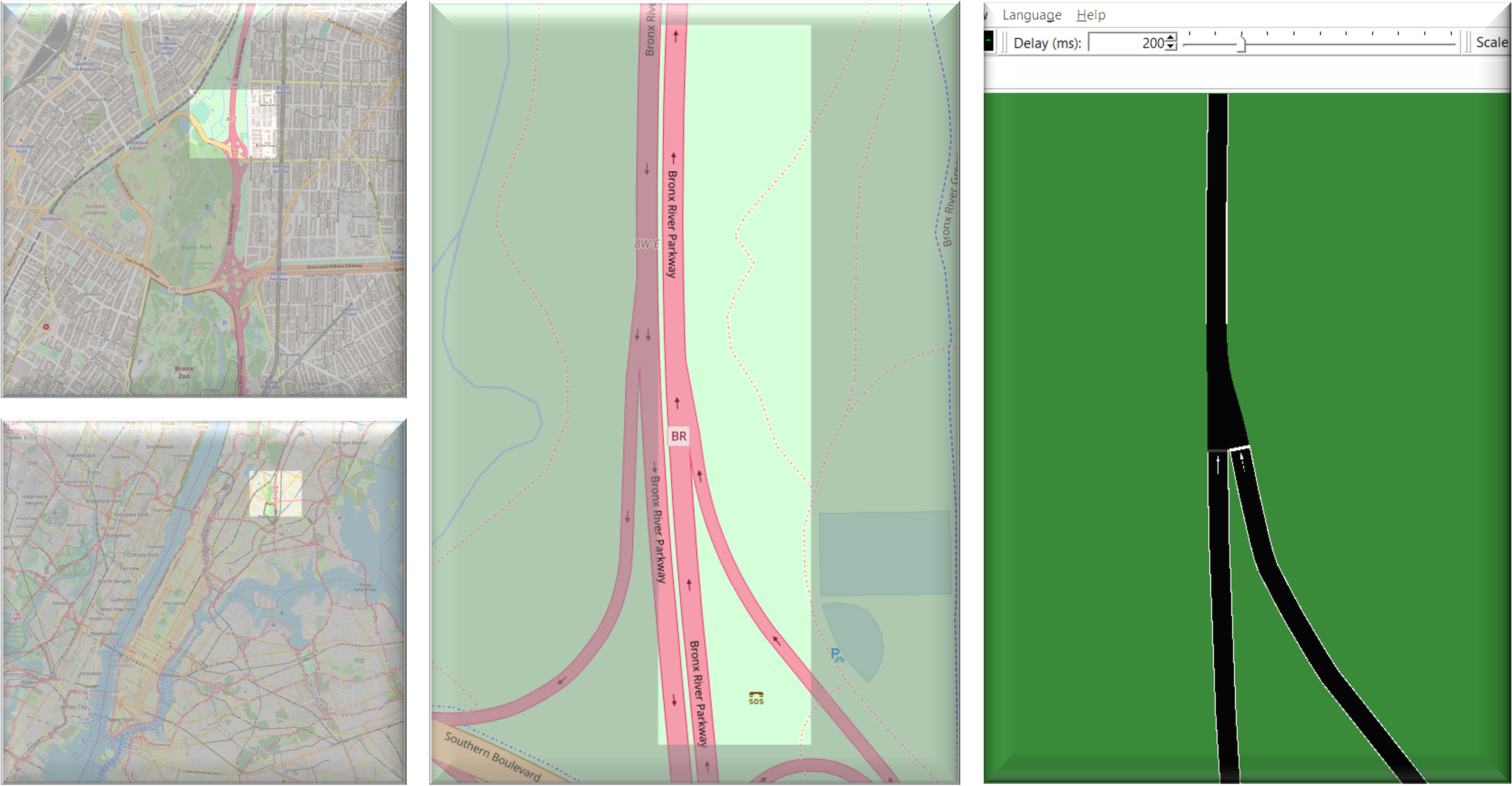}
\caption{Map of the merging on-ramp.}
\label{fig:map}
\end{figure}

Our study employed both MATLAB and SUMO (Simulation of Urban MObility) software along with the Traffic Control Interface (TraCI). TraCi allowed us to interfere with the simulation at each sampling step, enabling the vehicles to apply their feedback laws effectively. We focused on Junction 8E of the Bronx River Parkway, located several miles northeast of New York City. The location of this junction is illustrated in FIGURE \ref{fig:map}.  We transferred this specific junction to the SUMO software using data obtained from \cite{OpenStreetMap}. 

In our study, we assume that the initial distances between vehicles on the main road are sufficiently large (i.e., $>$50 m) for the following reasons: a) we clearly observe the change in acceleration only associated with the desired speed without any interaction of the repulsive forces generated by the function $V$, and b) although adopting a smaller inter-vehicle distance is feasible, a higher convergence point proves to be safer since it gives to the vehicles more time to react in case of an unexpected error

\subsection{Results}
\begin{figure*}[h!]
    \centering
    
    \begin{subfigure}[b]{0.48\textwidth}
        \includegraphics[width=\textwidth]{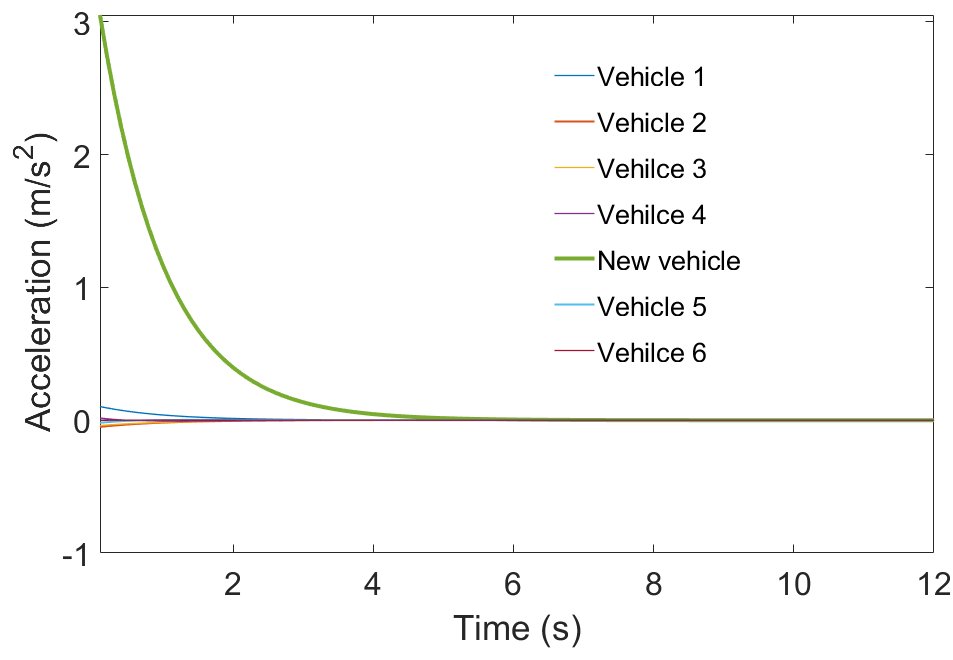}
        \caption{Acceleration with $\mu$ as in \cite{karafyllis2022stability}.}
        \label{fig:Acceleration1_old}

    \end{subfigure}
    \hspace{1.5mm}
    \begin{subfigure}[b]{0.48\textwidth}
        \includegraphics[width=\textwidth]{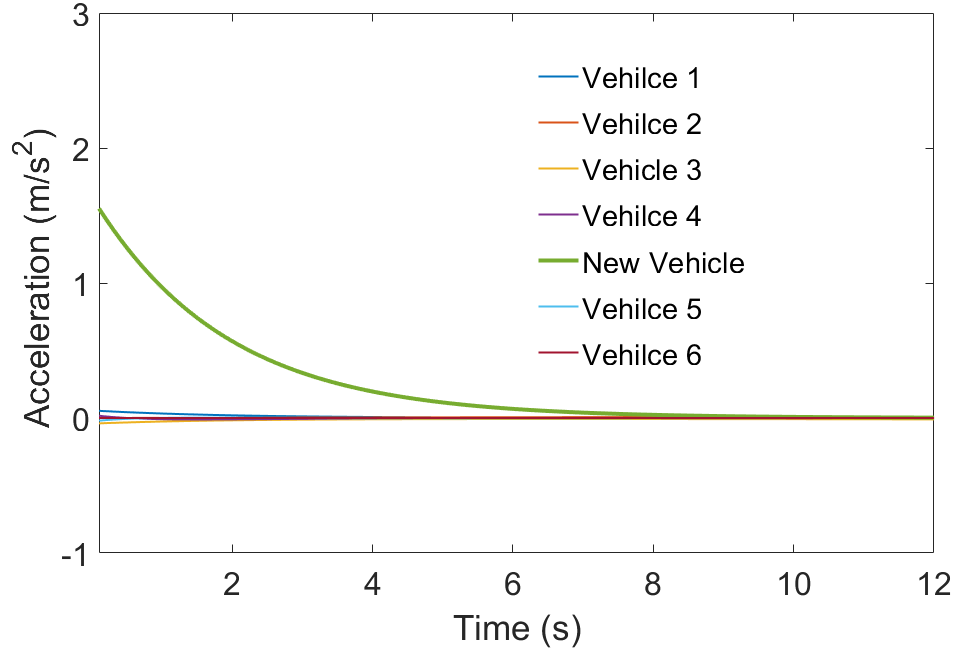}
        \caption{Acceleration using our new approach.}
        \label{fig:Acceleration1_new} 
    
    \end{subfigure}
        \caption{Acceleration after a new vehicle enters the road.}
    \label{fig:Accelerations_first_vehicle}

\end{figure*}

\begin{figure*}[h!]
    \centering
    
    \begin{subfigure}[b]{0.48\textwidth}
        \includegraphics[width=\textwidth]{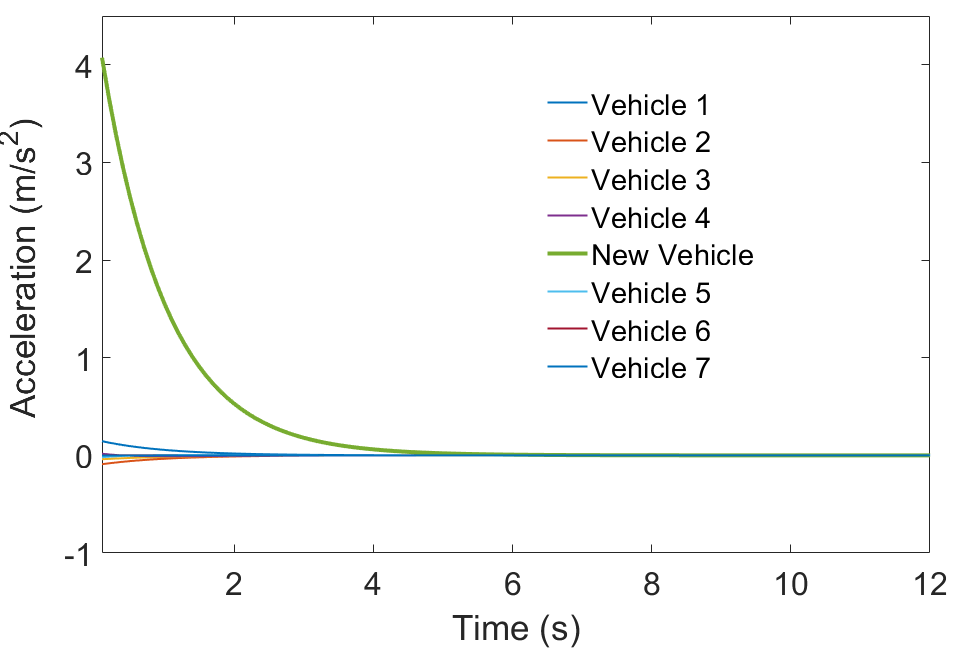}
        \caption{Acceleration with $\mu$ as in \cite{karafyllis2022stability}.}
        \label{fig:Acceleration2_old}

    \end{subfigure}
    \hspace{1.5mm}
    \begin{subfigure}[b]{0.48\textwidth}
        \includegraphics[width=\textwidth]{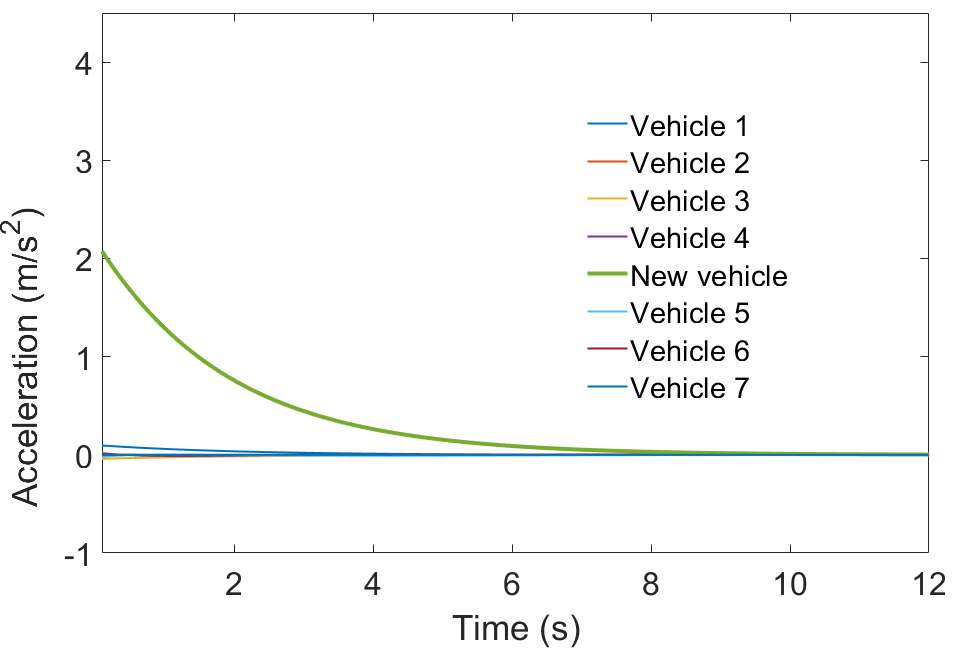}
        \caption{Acceleration using our new approach.}
        \label{fig:Acceleration2_new} 
    
    \end{subfigure}
        \caption{Acceleration after a second vehicle enters the road.}
    \label{fig:Accelerations_2nd vehicle}

\end{figure*}

In FIGURES \ref{fig:Accelerations_first_vehicle} and \ref{fig:Accelerations_2nd vehicle}, we see the accelerations in our system when two different vehicles merge to the main road, and we compare our results with the approach presented in \cite{karafyllis2022stability}. Inspecting FIGURES \ref{fig:Acceleration1_old} and \ref{fig:Acceleration1_new} we verify again that our approach yields the incoming vehicles to obtain smaller accelerations and slower converge to the desired speed. More specifically, we observe in FIGURE \ref{fig:Acceleration1_old} that when the new vehicle merged into the main road, it reached acceleration values around 3.05 m/s². On the other hand, as shown in Fig. \ref{fig:Acceleration1_new}, our approach reduces initial acceleration by approximately 50$\%$, down to 1.52 m/s². In FIGURES \ref{fig:Acceleration2_new} and \ref{fig:Acceleration2_old}, we confirm the same behaviour in our system when a second vehicle merges into the main road. Video of our SUMO simulation is provided through the following link:\\ https://sites.google.com/udel.edu/an-approach-for-optimizing/home.

\section{Concluding Remarks}
In this paper, we provided an approach that can aim at deriving the optimal values of a controller's gain proposed in \cite{karafyllis2022stability}, which significantly affects vehicle acceleration performance. Recognizing that vehicle dynamics constitute an initial value problem, we found that distinct initial conditions necessitate different optimal gain values. We framed and numerically solved an optimization problem to determine these optimal values. Due to the real-time infeasibility of solving this optimization problem, we trained a neural network that successfully learned the correlation between the initial conditions and the optimal gain value. Our approach reduced acceleration while being real-time implementable. A practical demonstration of our approach was provided through a vehicle merging scenario, emphasizing the significance of optimizing the gain based on different initial conditions. Ongoing research focuses on finding appropriate conditions that guarantee bounded acceleration and extend the same approach to lane-free roads.  
\section{Acknowledgements}
This research was supported by NSF under Grants CNS-2149520 and CMMI-2219761.

\section{Author Contributions}
The authors confirm their contribution to the paper as follows: study conception and design: Filippos N. Tzortzoglou; simulation: Filippos N.Tzortzoglou ;
analysis and interpretation of results: Filippos N. Tzortzoglou, Dionysios Theodosis, Andreas Malikopoulos; draft manuscript preparation: Filippos N. Tzortzoglou. All
authors reviewed the results and approved the final version of the manuscript.
\\

\bibliographystyle{unsrt}
\bibliography{trb_template,IDS_Publications_06112022}

\end{document}